\documentclass{scrartcl}

\usepackage{ae,aecompl,amsbsy,amssymb}
\usepackage{amsmath}
\usepackage{amsfonts}
\usepackage{mathpartir}
\usepackage{listings}
\usepackage{graphicx}
\usepackage{url}

\begin{document}
\title{On Behavioral Types for OSGi \\ \Large From Theory to Implementation}
\author{
Jan Olaf Blech$^{1,2}$ ~~~~ Harald Rue{\ss}$^1$  ~~~~ Bernhard Sch{\"a}tz$^1$}
\date{ \small $^1$fortiss GmbH, Munich ~~~~ $^2$RMIT University, Melbourne}

\maketitle

\begin{abstract}
This report presents our work on behavioral types for OSGi component systems. It extends previously published work \cite{reportosgisem,osgigeneraltech,isolavision12,umlfm} and presents features and details that have not yet been published. In particular, we cover a discussion on behavioral types in general, and Eclipse based implementation work on behavioral types . The implementation work covers: editors, means for comparing types at development and runtime, a tool connection to resolve incompatibilities, and an AspectJ based infrastructure to ensure behavioral type correctness at runtime of a system. Furthermore, the implementation comprises various auxiliary operations. We present some evaluation work based on examples.  
\end{abstract}

\section{Introduction}

In this report, we are extending the basic typing concepts of traditional software component systems with means for specifying possible behavior of components. As with traditional types, like primitive datatypes and their composition, our {\it behavioral types} can be used for eliminating possible sources of errors at development time of software systems. This is analog to classical static type checks performed by a compiler. Furthermore, we can use behavioral types for eliminating possible sources of errors at runtime. This is analog to dynamic type checks performed when accessing pointers that reference data with types that can not be statically determined in some classical programming languages. Behavioral types also provide additional information about components which can be used for tool based operations that allow the discovery of components and the dynamic reconfiguration of systems. We are focusing on the OSGi component framework \cite{osgi}.

The following topics are covered and have been partially  \cite{reportosgisem,osgigeneraltech,isolavision12,umlfm} published before: 
\begin{itemize}
\item
A discussion on behavioral types in general, including different usages.
\item
Our Eclipse based implementation work on behavioral types that is manifested in the {\sf BehT} framework. The implementation work covers: editors, means for comparing types at development and runtime, a tool connection to resolve incompatibilities, and an AspectJ based infrastructure to ensure behavioral type correctness at runtime of a system. Furthermore, the implementation comprises various auxiliary operations. 
\item
We present some evaluation work based on examples.  
\end{itemize}

\subsection{Core Concepts of Behavioral Types}
We present some core concepts on behavioral types to support a  development process of component based systems. In our opinion (behavioral) types should provide a number of core concepts \cite{umlfm} to justify their classification as a {\it type system}:
\begin{description}
\item[Abstraction: ] Behavioral types represent aspects of (models of) programs, components, or systems, providing an \emph{abstraction from details concerning the interaction with their environment as well as their internal structure}.  
\item[Type conformance: ] As -- in model-based development -- behavioral types are abstractions of components, models, or other entities. Type conformance is used to \emph{correctly relate a component to its behavioral type}. 
\item[Type refinement: ] For supporting stepwise refinement, behavioral types should  provide the concept of refinement to \emph{ensure the correct implementation of abstract specifications by concrete components}. 
\item[Type compatibility: ] For supporting the combination of components, behavioral types should provide the concept of type compatibility to \emph{help ensure the useful composition of components to systems.} 
\item[Type inference: ] Furthermore, for the same reason, behavioral types should provide the concept of type inference to \emph{allow to infer the type of a composed system from the types of its constituents.} 
\end{description}
To be useful in a development process, of course, a suitable type conformance notion has to be selected with respect to type refinement:  For a pair of models conforming  to a pair of types with the second model implementing the first, the second type should be a refinement of the first.  
Furthermore, type refinement, type compatibility and type inference should agree: If a type compatible to a given type is refined by another, the later type should be compatible to the given one; similarly, if in a composed type one type is replaced by a more refined type, the inferred type of the first composition should be a refinement of the second composition.  
Also, for practical application in a development process, a behavioral type should not only be explicitly provided for a component by the user and checked for conformance, but (automatically) constructed for this component. This is especially desirable in a seamless model-based development process.  
Finally, as type checking of expressive behavioral types is in general undecidable, an adequate level of expressiveness is needed making type checking feasible without over-restricting the expressiveness of the behavioral types.

\subsection{A Motivation for the Use of Behavioral Types}\label{subsec:Usage}

Using the above concepts, behavioral types can be helpful for different aspects of in the development process \cite{umlfm} of a component based system. Here, we present a general motivation for the concept without speaking about our implemented system.

\begin{figure}
\begin{center}
	\includegraphics[width=0.65\linewidth]{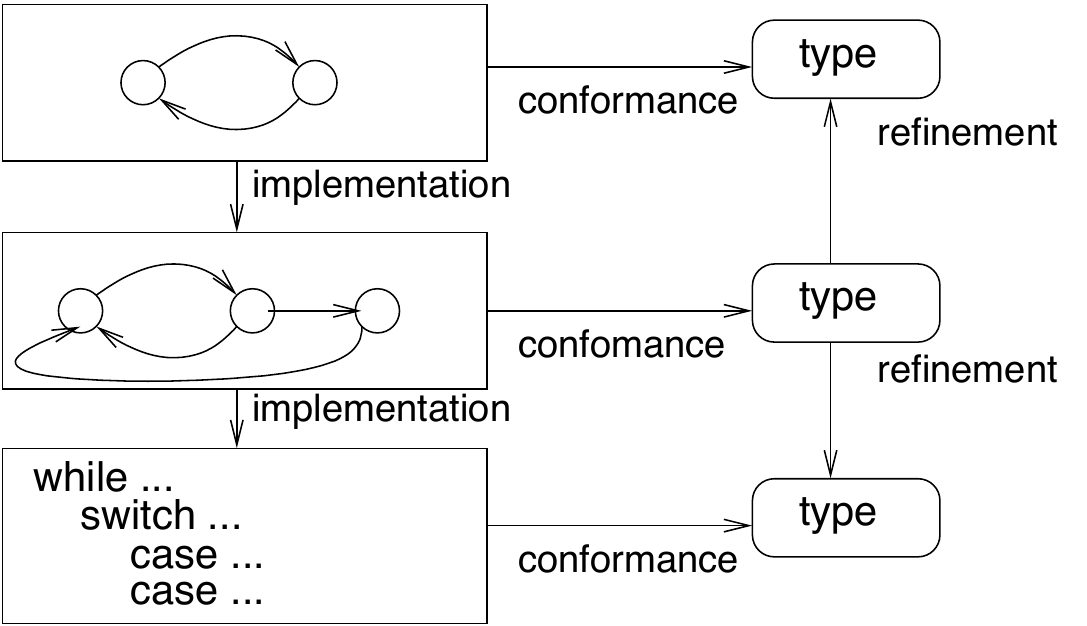}
\end{center}
	\caption{Refinement checking scenario}
	\label{fig:scenrefine}
\end{figure}

\begin{figure}
\begin{center}
\includegraphics[width=0.65\linewidth]{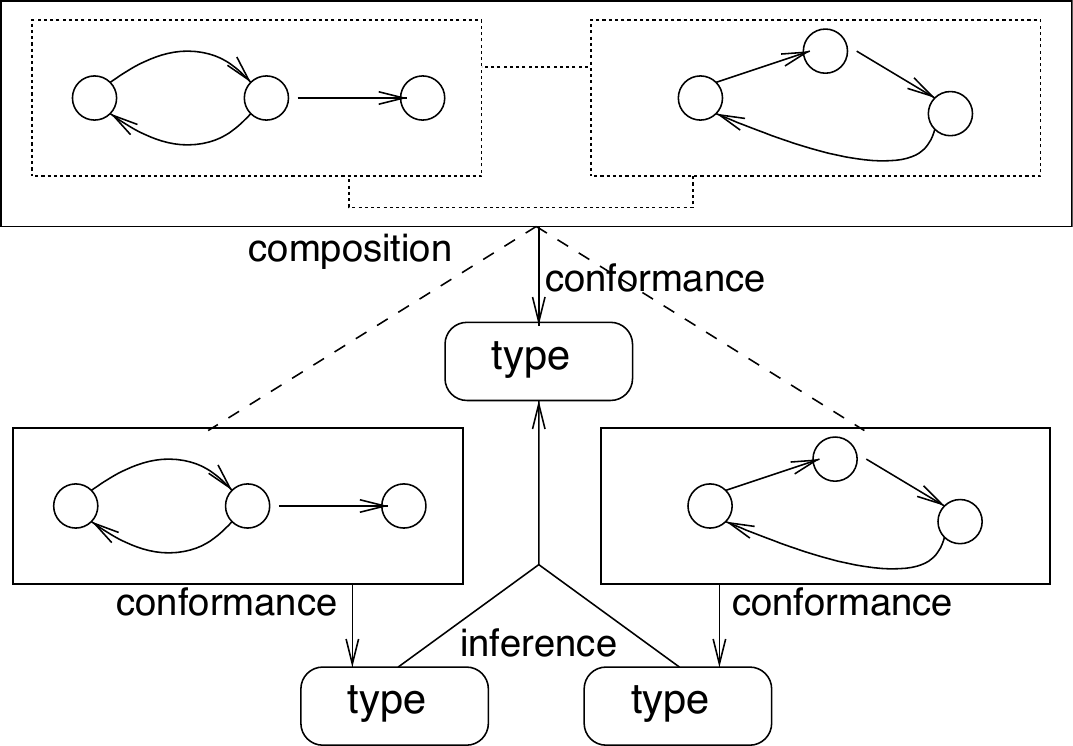}
\caption{Compound components scenario}
\label{Fig:compounds}
\end{center}
\end{figure}

\paragraph{Correctness of Implementation}
Behavioral types can be used to relate specifications and code, e.g., as products of different stages in a development process, to ensure a certain aspect of behavior is preserved. 
Figure~\ref{fig:scenrefine} illustrates this for a model based development process with models of different degrees of abstraction -- state-machines and source code --  representing the same system. Explicitly providing or automatically constructing  corresponding conforming types, correctness of refinement can be checked by using these types, ensure the correctness of the implementation with respect to the abstraction implied by the type system.
Furthermore, refinement checking is also used in structural refinement when implementing a component by a collection of subcomponents. As shown in Figure \ref{Fig:compounds}, the refinement relation is checked between the types of the composed components and the type of the collection of sub-components. The type of the composed component can also be derived from the respective behavioral types of the sub-components.

\begin{figure}
\begin{center}
	\includegraphics[width=0.65\linewidth]{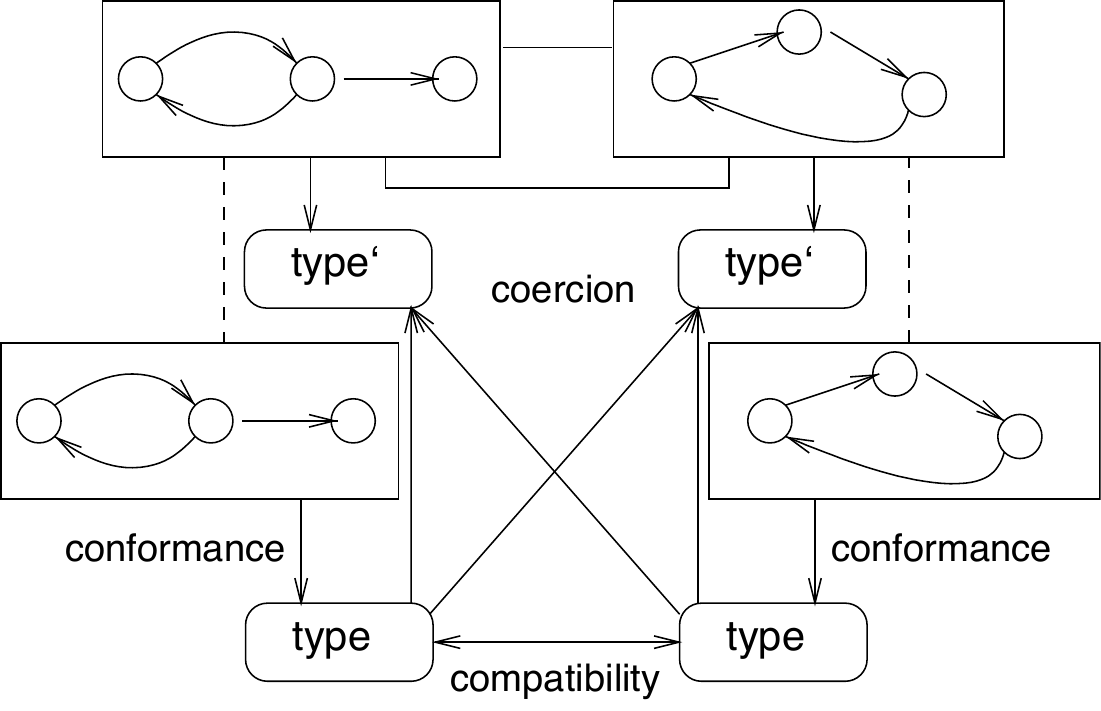}
	\caption{Compositionality checking scenario}
	\label{fig:scencomprefine}
\end{center}
\end{figure}

\paragraph{Compositionality and Interfaces}
 Behavioral types could furthermore be used to check whether components to be composed are compatible with each other, 
as shown in Figure~\ref{fig:scencomprefine}.
Additionally, using type coercion -- i.e., the inference of the least abstract types refining the investigated pairs and being compatible -- basically incompatible components can be composed. This generally involves an adaption of the corresponding models -- by providing ``glue code'' similar to automatic type casts for, e.g., integers of different length -- to make the two components interact with each other. 

\begin{figure}
\begin{center}
	\includegraphics[width=2.6in]{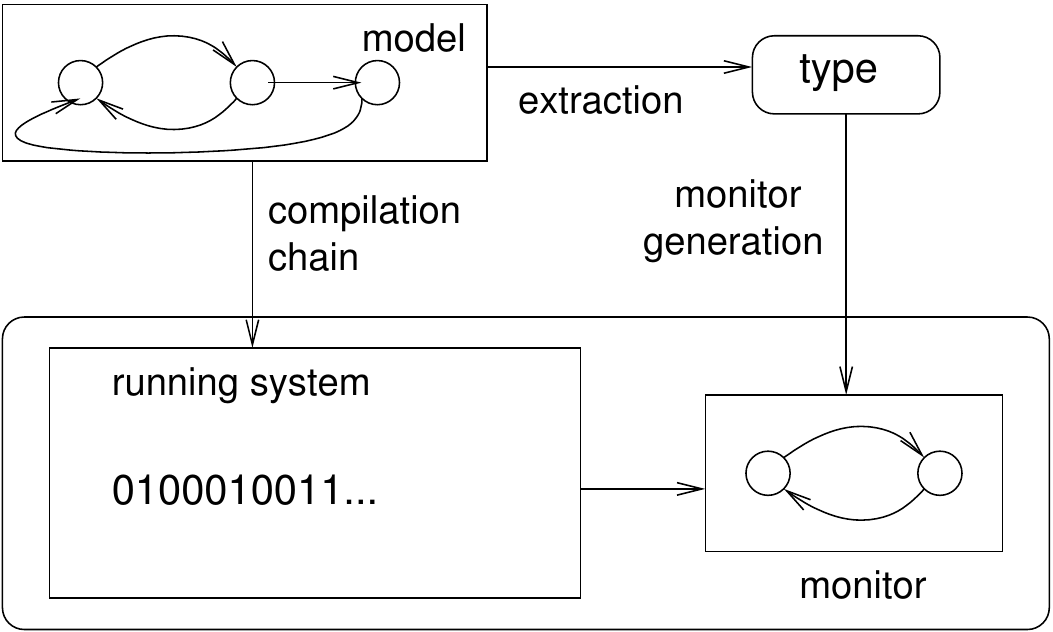}
	\caption{Behavioral types as monitors in RV}
	\label{fig:scenRV}
\end{center}
\end{figure}

\paragraph{Runtime Verification}
Finally, behavioral types can be used for runtime-verifying systems, supplying a monitor being executed in parallel with a system implementation. The monitor -- corresponding to a behavioral type and checking all behavioral constraints specified via the type -- observes the system behavior, reporting violations. Monitors may be generated from behavioral types automatically.
Figure~\ref{fig:scenRV} shows a usage scenario for behavioral types in runtime verification. Using an explicitly provided type or inferring it from the specification (a more abstract model in the model based development process shown), this type serves as the basis for generating a runtime monitor which is then deployed with the compiled model as a runtime-verified system. This aspect is not treated in our current implementation.

\paragraph{Additional Benefits}
Additional benefits comprise the dynamic reconfiguration of systems based on type information and the discovery of components in a SOA like setting.

\subsection{An Example for Refinement of Behavioral Types}

\begin{figure} 
\begin{center}
	\includegraphics[width=3in]{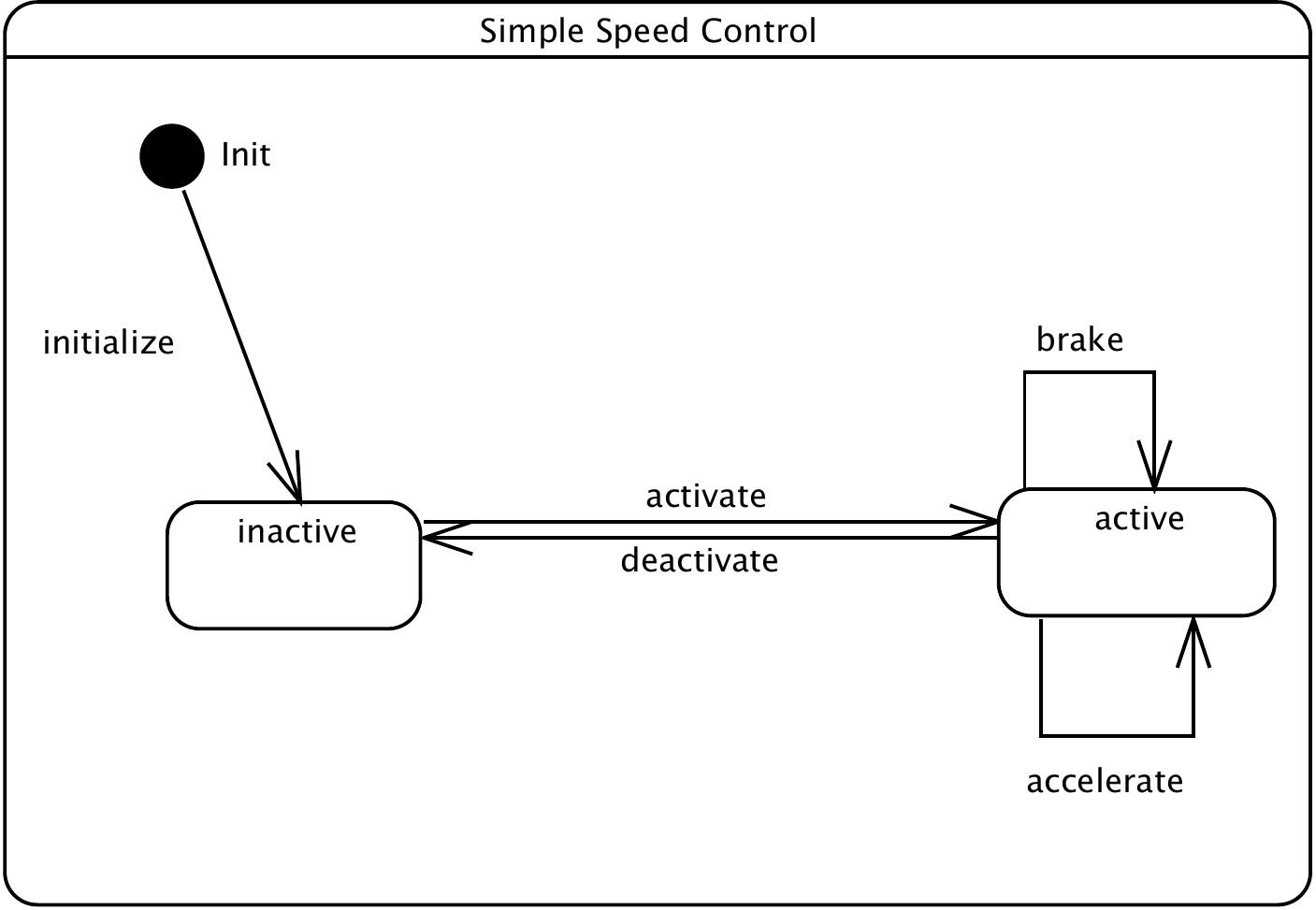}
\end{center}
\caption{A simple speed control state-machine}
\label{fig:sscpic}
\end{figure} 

\begin{figure*} 
\begin{center}
	\includegraphics[width=4in]{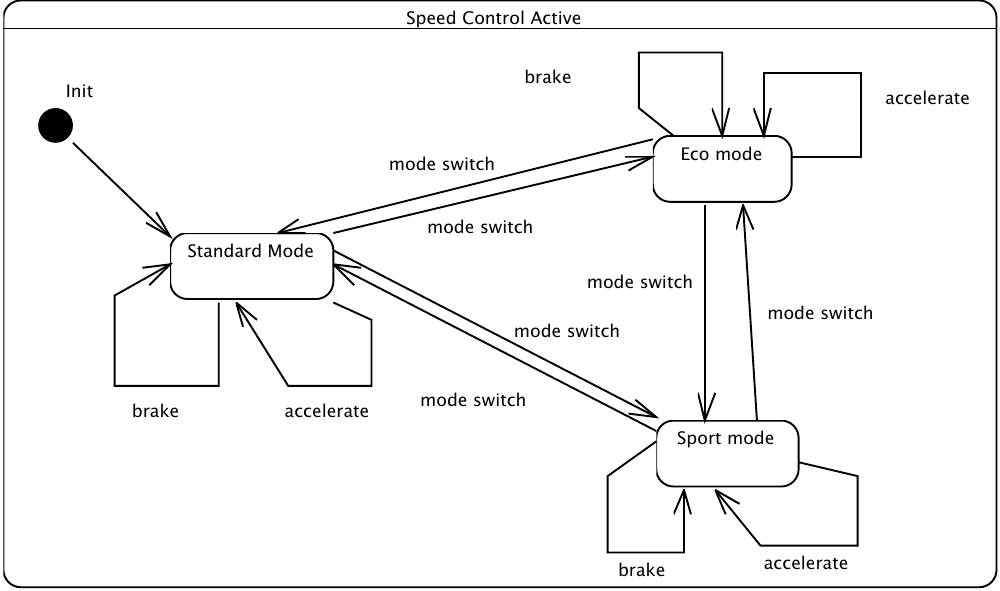}
\end{center}
\caption{Refining the active state}
\label{fig:ascpic}
\end{figure*}

Figure~\ref{fig:sscpic} shows a state-machine  ''speed control'' as part of an adaptive cruise control system in a car, using the graphical model of the Eclipse-based Papyrus UML tool \cite{papyrus}. It is taken from \cite{umlfm}. This state-machine provides an abstract component specification created during requirements specification in the development process. It specifies that this component shall be able to perform acceleration and braking. 
It can be compared with other types for UML diagrams that specify some aspects of the behavior of ``speed control''. This comparison can be used during the abstract and detailed design and -- not covered in this paper -- in a later implementation, supporting a stepwise refinement. For example in the next phase the active state can be specified in a more detailed way supporting several modes as shown in Figure~\ref{fig:ascpic}. The standard, eco and sport mode may show different  acceleration and braking behavior thereby supporting, e.g., more fuel-efficient driving in the eco mode.  
However, when abstracting from possible transition guards (mode switch), other behavioral functionality and other events -- these may limit the order of possible executions -- the original behavior specification still applies: Each mode supports braking and acceleration. We can now extract a behavioral type of this more refined model and compare it with the first one.  
On this abstraction level -- regarding only  brake and acceleration guards -- both specifications have the same set of execution traces. 

 As an ultimate goal, the development environment should support the extraction and checking of behavior  automatically and provide a means of informing the developer about any behavioral incompatibilities, i.e., understandable behavioral type errors. 

\subsection{OSGi}
\label{sec:osgi}
\begin{figure}[t]
\centering
\includegraphics[scale=0.75]{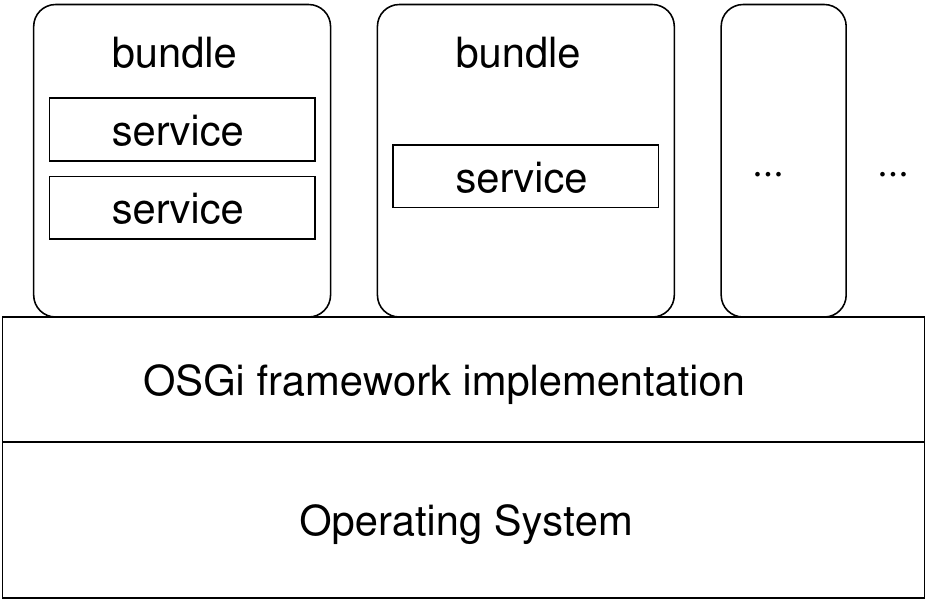}
\caption{OSGi framework}
\label{fig:osgiexample}
\end{figure}

We present an overview on OSGi following our description in \cite{isolavision12} and refer to our semantics report \cite{reportosgisem} for our approach to cover the semantics of OSGi (parts of this has  also been published in \cite{osgigeneraltech}).

The OSGi framework is a component and service platform for Java. It allows the aggregation of Java packages and classes into bundles (cf. Figure~\ref{fig:osgiexample}) and comes with additional deployment information. The deployment information  triggers the registration of services for the OSGi framework.
Bundles provide means for dynamically configuring services, their dependencies and usages. OSGi bundles are used as the basis for Eclipse plugins but also for embedded applications including solutions for the automotive domain, home automation and industrial automation. Bundles can be installed and uninstalled during the runtime. For example, they can be replaced by newer versions. Hence, possible interactions between bundles can in general not be determined statically.

Bundles are deployed as .jar files containing extra OSGi information. This extra information is stored in a special file inside the .jar file. Bundles generally contain a class implementing an OSGi interface that contains code for managing the bundle, e.g., code that is executed upon activation and stopping of the bundle. 
Upon activation, a bundle can register its services to the OSGi framework and make it available for use by other bundles. Services are implemented in Java. The bundle may itself start to use existing services. Services can be found using dictionary-like mechanisms provided by the OSGi framework. Typically one can search for a service which is provided using an object with a specified Java interface. 

In the context of this report, we use the term OSGi component as a subordinate concept for bundles, objects and services provided by bundles.

The OSGi standard only specifies the framework including the syntactical format specifying what bundles should contain. Different implementations exist for different application domains like Equinox\footnote{\url{http://www.eclipse.org/equinox/}} for Eclipse, Apache Felix\footnote{\url{http://felix.apache.org/site/index.html}} or Knopflerfish\footnote{\url{http://www.knopflerfish.org/}}. If bundles do not depend on implementation specific features, OSGi bundles can  run on different implementations of the OSGi framework.

\subsection{Overview}
Section~\ref{sec:beht} discusses and presents a general work on behavioral types. The use of behavioral types an the development of OSGi components is described in Section~\ref{sec:osgidevproc}. BehT, our tool is discussed in Section~\ref{sec:tool} together with related implementation questions. An evaluation is described in Section~\ref{sec:eval}. Related work is discussed in Section~\ref{sec:rw} and a conclusion featured in Section~\ref{sec:concl}.

\section{Behavioral Types}
\label{sec:beht}

Here, we present and discuss a general implementation independent concept of behavioral types.

\subsection{Finite Automata and Regular Expression Based Behavioral Types}

Our behavioral types essentially support finite automata and regular expressions as the main specification format. Finite automata and regular expressions can easily be transformed in one another. Finite automata are used for specifying expected incoming, potential outgoing method calls, the creation and deletion of components during a time span and other events that may occur in the lifetime of a system. A component's behavior can be specified by one or multiple automata each one describing a behavioral aspect. Formally, we have an alphabet of labels $\Sigma$, a set of locations $L$, an initial location $l_0$ and a set of transition edges $E$ where each transition is a tuple $(l,\sigma,l')$ with $l,l' \in L$ and $\sigma \in \Sigma$. These are aggregated into a tuple to form a behavioral specification: 
\begin{center}
$(\Sigma, L,l_0,E)$
\end{center}
This view abstracts from the specifications given in Section~\ref{sec:osgi}. Our intention is to define interaction protocols or some aspects of them like the expected order of incoming and outgoing method calls for a component. Specifications for different components are independent of each other as long as there is no method call (e.g., indicated by the same label name) in the specifications.

\begin{figure}
\centering
\includegraphics[scale=0.85]{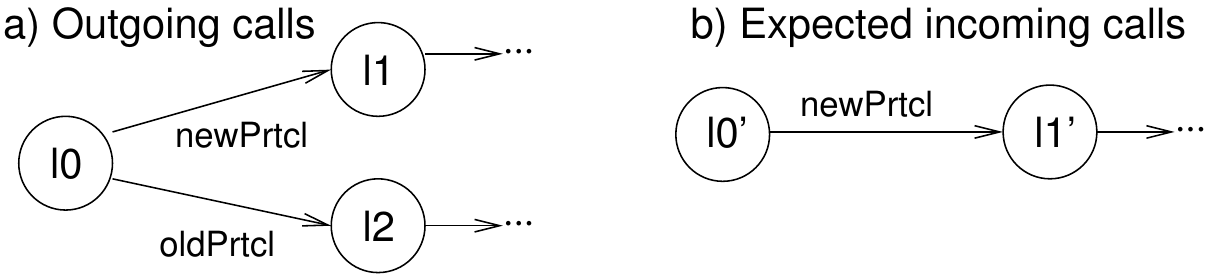}
\caption{Supporting different protocol versions}
\label{fig:oldnewprot}
\end{figure}

\paragraph{Example: Two components interacting}
Specifications can be used for different behavioral aspects. Figure~\ref{fig:oldnewprot} shows two excerpts of automata for outgoing and expected method calls from two different component specifications: 
\begin{center}
{\small $(\{newPrtcl, oldPrtcl, ... \}, \{l0,l1,l2,...\}, l0, \{(l0,newPrtcl,l1),(l0,oldPrtcl,l2) ,...\})$ \\
\vspace{0.2cm}
and \\ 
\vspace{0.2cm}
$(\{newPrtcl, ... \}, \{l0,l1,...\}, l0, \{(l0,newPrtcl,l1) ,...\})$ }
\end{center}
Here, the first component can do two different method calls in its initial state: {\sf newPrtcl, oldPrtcl}. The second component expects one method call {\sf newPrtcl} in its initial state. 
In this case both components may interact with each other, if both components use the {\sf newPrtcl}. 
\paragraph{Interaction protocols for bundles and objects}

Objects and bundles can register a service protocol -- describing, e.g., incoming method calls -- that they expect. This can be done by using:
\begin{itemize}
\item {\it Regular expressions.} Thereby bundles and objects can indicate expected events. Events can be incoming or outgoing method calls. Thus, the regular expression specifies their order. Regular expressions are terms over an alphabet of events using the $+$ for alternatives, the $.$ for concatenation and the $^*$ as the star operator.
\item {\it Finite automata.} Regular expression can be described by an equivalent finite automaton, too. We define our finite automata as a set of locations, an initial location and a transition relation comprising a predecessor and a successor location labeled with an event.
\end{itemize}
While in our applications the event is typically a method call or a set of method calls, other possibilities like timing events, or creation and deletion of objects and bundles are also possible.
 
\begin{figure}
\centering
\includegraphics[scale=0.75]{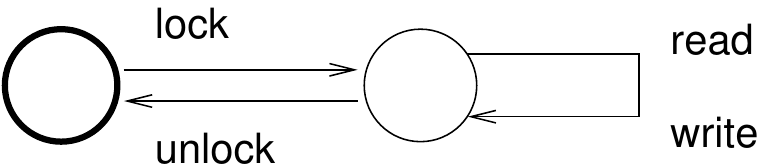}
\caption{Intended interplay of method calls}
\label{fig:osgiservicefile2}
\end{figure}

For example the protocol given in Figure~\ref{fig:osgiservicefile2} can be described as a regular expression as follows:
\begin{center}
((INC: Lock)$.$(INC: Read $+$ INC: Write)$^*$$.$(INC: Unlock))$^*$
\end{center}
The expression describes a sequence that can be repeated. It starts with a lock and ends with an unlock. Between lock and unlock an arbitrary number of read and write operations can occur. The INC denotes expected incoming method call. 

The actions from  Figure~\ref{fig:osgiservicefile2} describe outgoing method calls. This can be written using our regular expressions as:
\begin{center}
(OUT: LockF1)$.$(OUT: LockF2)$.$ \\
((OUT: ReadF1) $+$ (OUT: ReadF2) $+$ (OUT: WriteF1) $+$ (OUT: WriteF2))$^*$$.$ \\
(OUT: UnlockF2)$.$(OUT: UnlockF1) 
\end{center}
and
\begin{center}
(OUT: LockF2)$.$(OUT: LockF1)$.$ \\
((OUT: ReadF1) $+$ (OUT: ReadF2) $+$ (OUT: WriteF1) $+$ (OUT: WriteF2))$^*$$.$ \\
(OUT: UnlockF1)$.$(OUT: UnlockF2) 
\end{center}
One can now use these protocol specifications, e.g., for checking:
\begin{itemize}
\item {\it Compatibility} This addresses the question if  the operations that one object expects to be called are called by another object. Furthermore, the correct order of calls is of interest.
\item {\it Additional properties} Properties that relate distinct semantical aspects of bundles and objects are of interest. In the given example, the question arises whether a deadlock can occur or not.
\end{itemize}
In order to perform these checks and analysis one has to match elements of a specification for one component with elements of a specification from another component. In the given example the protocol comparison has to deal with two instances of a file component and has -- for example -- to relate the (OUT: LockF1) and (OUT: LockF2) with instances of (INC: Lock).

\paragraph{Parameterized specifications}
For facilitating the relation of specifications we define parameterized specifications. These comprise:
\begin{itemize}
\item {\it Parameterized regular expressions.} Here, each event used in a regular expression can be augmented with a parameter. For our example file component specification this results in the following expression, parameterized with $<F>$.
\begin{center}
((INC: Lock$<F>$)$.$(INC: Read$<F>$ $+$ INC: Write$<F>$)$^*$$.$(INC: Unlock$<F>$))$^*$
\end{center}
\item
{\it Parameterized automata.} Similar to regular expressions, locations and events in transitions of automata can be augmented with parameters.
\end{itemize}
Instantiation is done, by substituting concrete values for the parameter. Instantiation of parameters is dependent on concrete application scenarios. 
\paragraph{Example instantiations of parameterized specifications}
We regard two kinds of instantiations as particularly useful for describing a protocol of expected incoming method calls.

\begin{figure}
\centering
\includegraphics[scale=0.75]{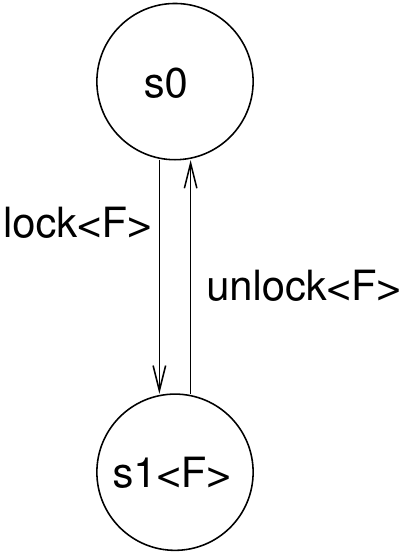}
\caption{Parameterized specification for locking / unlocking}
\label{fig:instparamspec}
\end{figure}

Consider the refined version of Figure~\ref{fig:osgiservicefile2} in Figure~\ref{fig:instparamspec} for locking and unlocking a resource. The lock state as well as the method calls that lead to the lock state are parameterized.

\begin{itemize}
\item A first instantiation is shown in Figure~\ref{fig:instparam}. Here, the parameter is instantiated by instances $f0,...,fn$. Each of them gets its own lock state and its own method call that lead to this lock state.
\begin{figure}
\centering
\includegraphics[scale=0.75]{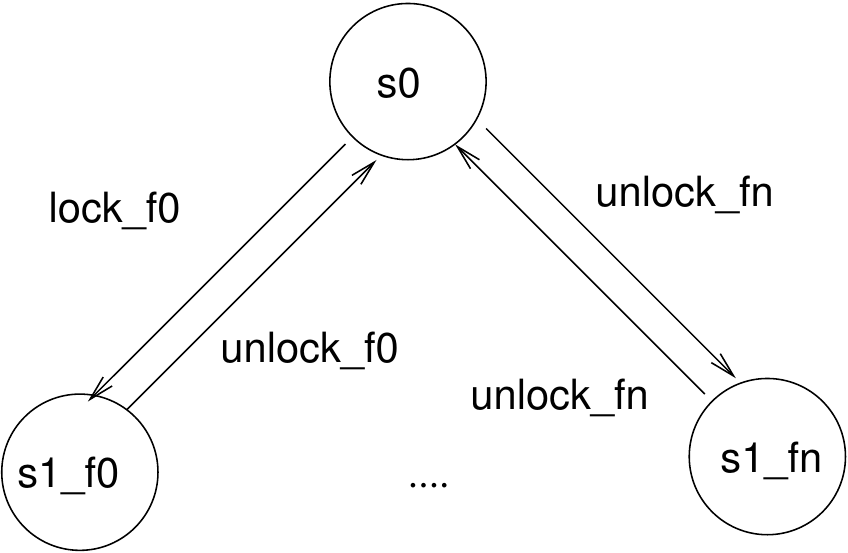}
\caption{Instantiating a parameterized specification}
\label{fig:instparam}
\end{figure}

\item In case only one lock state is wanted, one can still deal with different parameterized method calls and use the instantiation shown in Figure~\ref{fig:instparamex}
\begin{figure}
\centering
\includegraphics[scale=0.75]{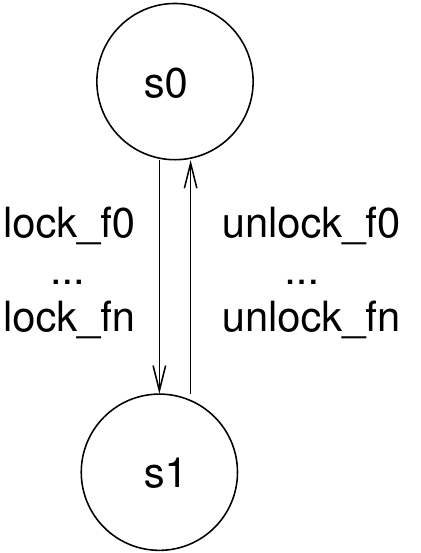}
\caption{Another way of instantiating a parameterized specification}
\label{fig:instparamex}
\end{figure}
\bibliographystyle{plain}
\end{itemize}

\section{Behavioral Types at Development and Runtime of a System for OSGi}
\label{sec:osgidevproc}

A potential major advantage by using  behavioral types is the support of a seamless integration of behavioral specification throughout the development phase and the life cycle of a system.  

\begin{figure}
\centering
\includegraphics[width=0.75\textwidth,angle=0]{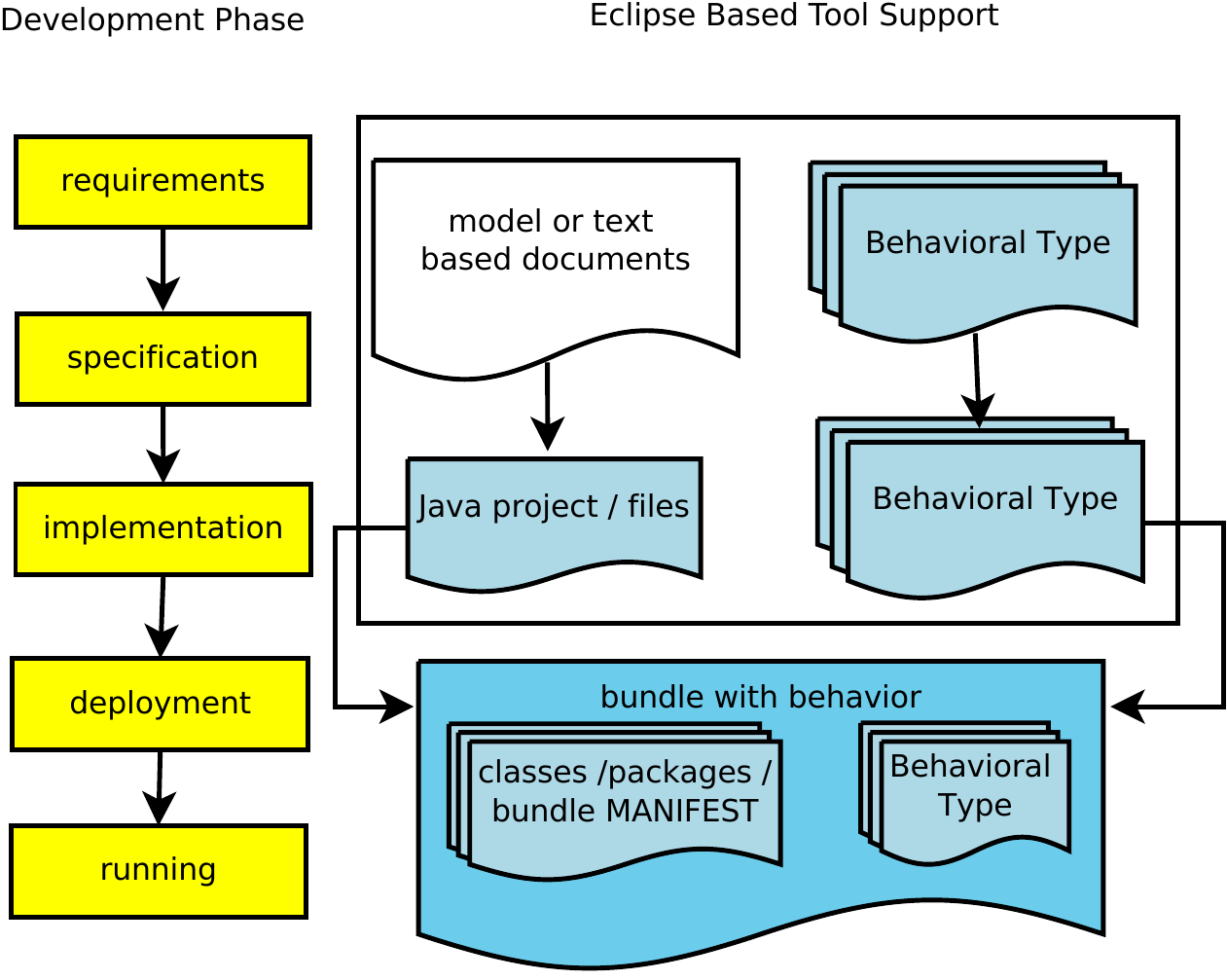}
\caption{Behavioral types at development time}
\label{fig:devchain}
\end{figure}

Our behavioral types can be used for different purposes (we proposed them partially in \cite{umlfm}) at development  and runtime.
The main idea of using behavioral types at development time is to derive them from requirements as shown in Figure~\ref{fig:devchain} and use them for 
\begin{itemize}
\item
refinement checking of different forms of specification for the same entity that are supposed to have some semantical meaning in common. For example, the abstract specification, source code and compiled code of the same component represent different abstraction levels and should fulfill the same behavioral type. Checking this could be done by using static analysis at development time. 
\end{itemize}
At the end, a developed OSGi bundle is deployed including the behavioral type files. These can now be used for additional (dynamic) operations in the running system.

A feature that can be performed at compile and at runtime is 
\begin{itemize}
\item the compatibility checking for the composition of software components and the generation of glue code -- for behavioral type coercion -- to overcome possible incompatibilities 
\end{itemize}

At runtime of a system this feature can help  dynamic reconfiguration.

\begin{figure}
\centering
\includegraphics[width=0.75\textwidth,angle=0]{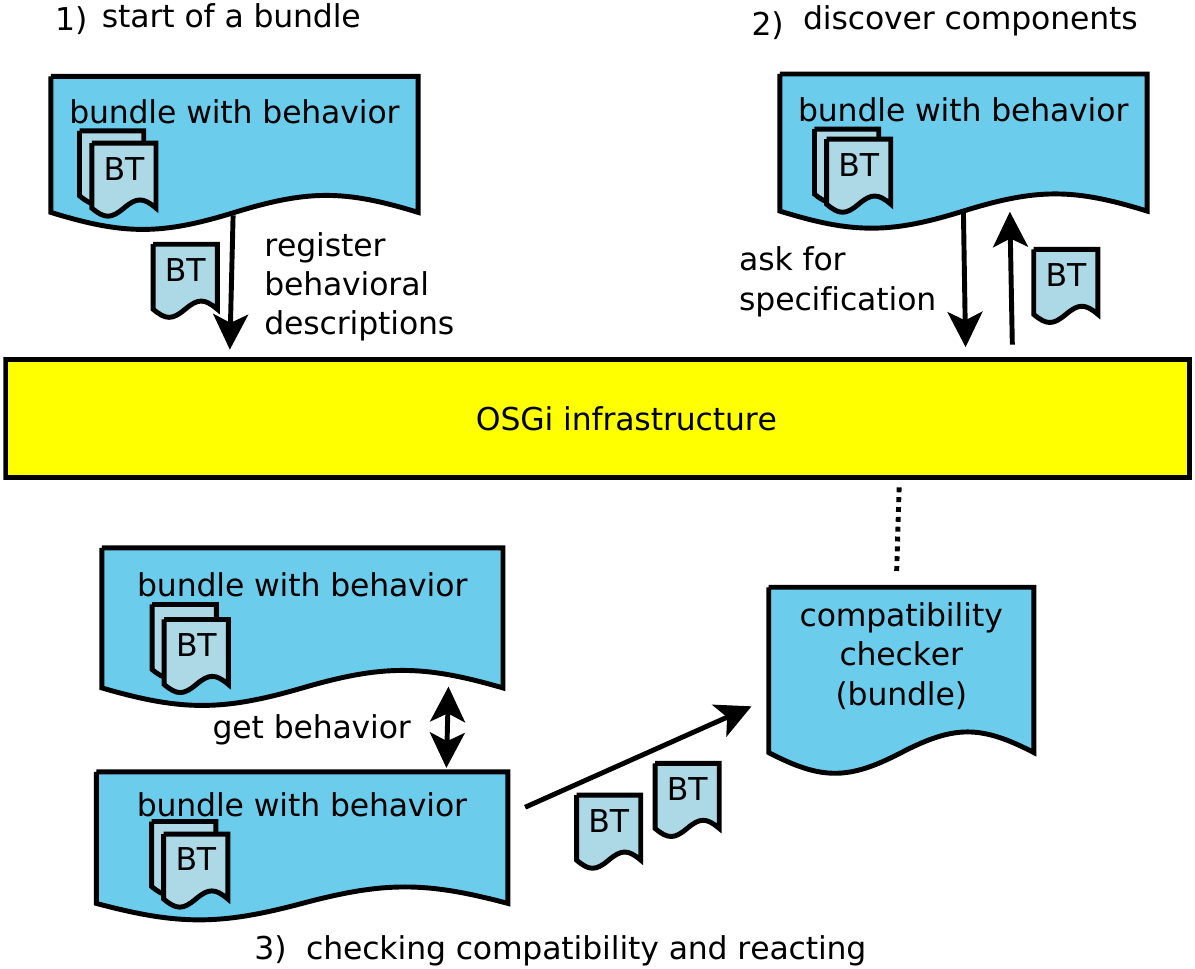}
\caption{Behavioral types at runtime}
\label{fig:behtrt}
\end{figure}

Figure~\ref{fig:behtrt} shows two operations which can be carried out at runtime of a system:
\begin{itemize}
\item the registration and discovery of components using the OSGi framework,
\item the compatibility, e.g., deadlock checking of bundle interaction protocols.
\end{itemize}

Furthermore, another operation that could be invoked at runtime is
\begin{itemize} 
\item  the adaptation of a component  to act according to a required protocol. This can be a solution for dynamic type coercion.
\end{itemize}

\begin{figure}
\centering
\includegraphics[width=0.75\textwidth,angle=0]{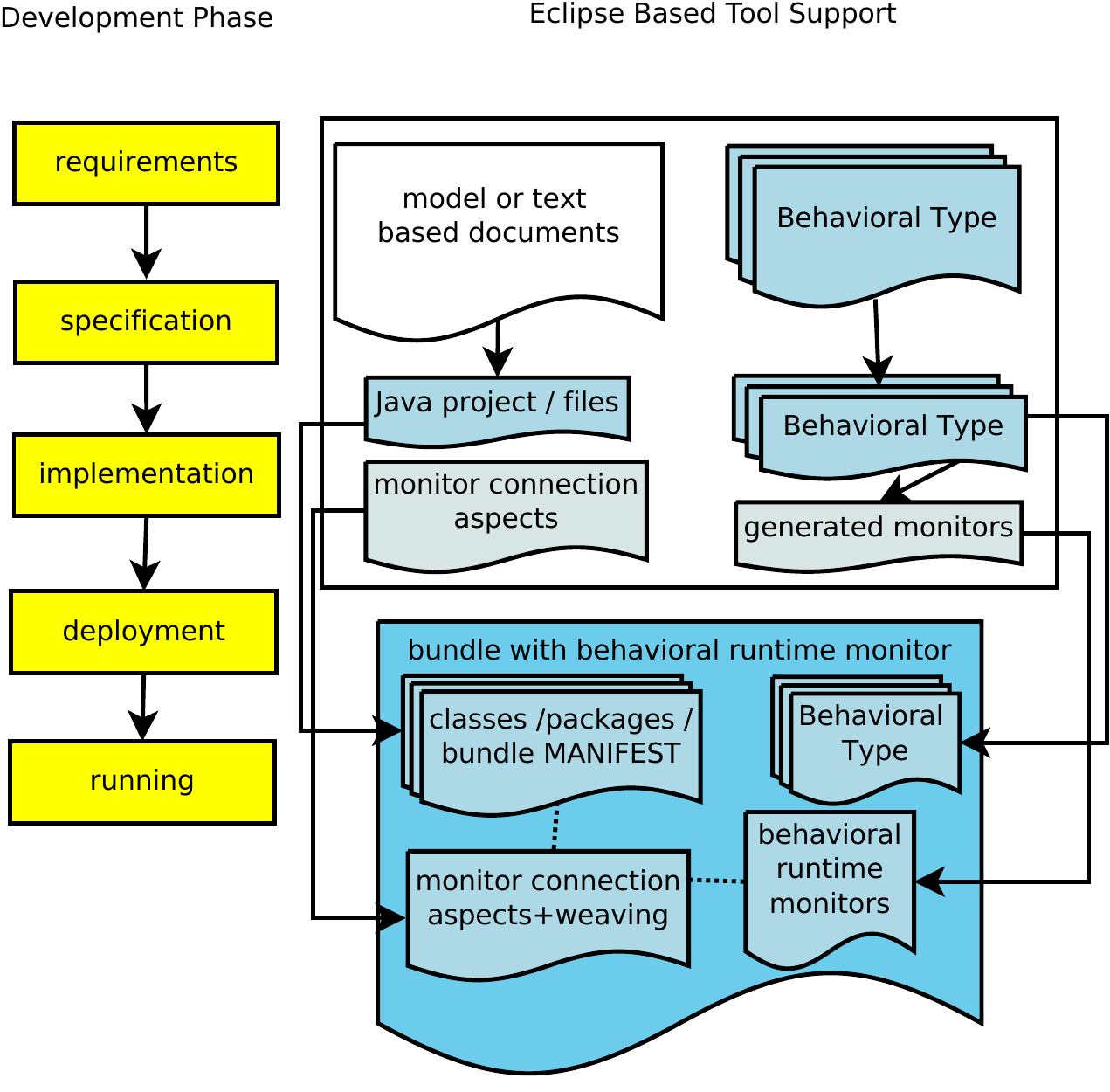}
\caption{Behavioral runtime monitors}
\label{fig:behtrvmon}
\end{figure}

Behavioral runtime monitors (Figure~\ref{fig:behtrvmon}) as featured in this paper comprise
\begin{itemize}
\item the generation of the behavioral runtime monitor and its connection using aspects at development time and
\item the actual monitoring at runtime.
\end{itemize}

\section{BehT: An Eclipse Based Framework for Behavioral Types}
\label{sec:tool}

This section presents our implementation work on BehT: the Eclipse based framework for behavioral types of OSGi components. Some parts are already published in \cite{osgigeneraltech}.

\subsection{Behavioral Types in BehT}

Our behavioral types provide an abstract description of a components behavior and thus provide a way of formalizing  specifications associated with the component. They can be used as a basis for checking the compatibility of components -- for composing components into new ones, and interaction of different components  -- and for providing ways to make components compatible using coercion. Type conformance can be enforced at compile time (e.g., like primitive datatypes int and float in a traditional typing system) -- if decidable and feasible -- or at runtime of a system -- e.g., like whether a pointer is assigned to an object of a desired type at runtime in a traditional typing system. 

In our work behavioral types are realized as files that contain a description of (parts of the) behavior of an OSGi component. Typically, there should be one file per bundle, or class definition. But different aspects of behavior may also be realized using different files. In Eclipse the files are associated with an OSGi bundle by putting them in the same project folder in the Eclipse workspace.

Here, behavioral types are formally defined using the following ingredients.
\paragraph{Behavioral Type Automaton}
A behavioral type automaton is a finite automaton represented as a tuple $(\Sigma, L,l_0,E)$ comprising an alphabet of labels $\Sigma$, a set of locations $L$, an initial location $l_0$ and a set of transition edges $E$ where each transition is a tuple $(l,\sigma,l')$ with $l,l' \in L$ and $\sigma \in \Sigma$.
A consistency condition on our types is that all $\sigma \in \Sigma$ appear in some transition in $E$.

In this paper, since we are interested in method calls, $E$ is the set of method names of components. The definition presented here can be used for specifying the behavior of single objects, all objects from a classes, bundles and their interactions. It can be used for monitoring incoming method calls, outgoing method calls, or both. 

\paragraph{Maximal Execution Time Table}
In addition to the protocol defined by the behavioral type automaton, we define the maximal execution time of methods as a mapping $\Sigma \rightarrow long \cup \bot$ from the set of method names $\Sigma$ to their maximal execution time in milliseconds. The specification of a maximal execution time is optional, thus, the $\bot$ entry indicates that no maximal execution time is set.

\paragraph{Behavioral Types}
A behavioral type in our framework may comprise a behavioral type automaton and a maximal execution table. In addition to this, it may comprise parameterized specifications, LTL formula, regular expressions and information on what is specified. Here, indications on the nature of events and textual descriptions are available.

\subsection{Behavioral Runtime Monitors from Behavioral Types}
We have implemented a behavioral runtime monitor generator as described in Section~\ref{sec:osgidevproc} for BehT following the outline of Figure~\ref{fig:behtrvmon}.

Regardless of what we intend to monitor, the monitor generation from a specification is the same. It is done automatically from a behavioral type file and generates a single Java file that defines a single monitor class.

Figure~\ref{fig:genmonex} shows a generated monitor. Monitors are generated as classes bearing a name derived from the original behavioral type. They comprise a map {\tt maxtimes} that maps method names to their maximal execution time in milliseconds.  This entry is optional. If present, this map is initialized by the constructor --
\begin{center}
{\tt public clientinstance\_out\_realistic\_simple\_mon()} 
\end{center}
in the example -- of the monitor with the values specified for methods in the behavioral type file. Generated from an automaton from the behavioral type our behavioral runtime monitors comprise a static enumeration type with the location names of the automaton. In the automaton, the locations {\tt LOCs0, LOCs1} are present.   Using this type a state transition function generated from the transition relation is generated.  The state transition function takes a string encoding a method name -- event name -- and updates a state field {\tt protected LOCATION state} of the method. This field is initialized on object creation with the name of the initial state: {\tt LOCs0} in the example. 
\begin{figure}
\centering
{\small
\begin{verbatim}
package monitors;

import ....

public class clientinstance_out_realistic_simple_mon { 

   public Map<String,Long> maxtimes = new HashMap<String,Long>();

   public clientinstance_out_realistic_simple_mon() {
      maxtimes.put("listFlights",new Long(1000));
   }

   public static enum LOCATION { 
     LOCs0 , LOCs1
   }

   protected LOCATION state = LOCATION.LOCs0;

   public boolean nextState(String event) { 
      boolean rval = false;
      switch (state) {
         case LOCs0:
            if (event.equals("newMiddlewareProc")) {
               state = LOCATION.LOCs1;
               rval = true;
            }
            break;
         case LOCs1:
            if (event.equals("listFlights")) {
               state = LOCATION.LOCs1;
               rval = true;
            }
            
            ....

            if (event.equals("listFlight")) {
               state = LOCATION.LOCs1;
               rval = true;
            }
            break;
      }
      return rval;
   }
}
\end{verbatim} 
}

\caption{Generated example monitor}
\label{fig:genmonex}
\end{figure}

\subsection{Behavioral Runtime Monitor Integration using AspectJ}
\label{sec:aspects}
The generated monitors are connected to the component that shall be observed using AspectJ aspects. AspectJ is an extension of Java that features aspect oriented programming. Aspects are specified in separate files and feature pointcuts that allow the specification of locations where Java code specified in the aspect shall be added to existing Java code. This weaving of aspect code into existing Java code is done on bytecode level.  

Monitors are created and called from aspects. All extra code needed to integrate the monitors is defined in the AspectJ files or in libraries accessed through the AspectJ files. There is no need to touch the source code of a component. This independence of source code and specification is a design goal of our framework.
We distinguish different kinds of monitor deployment. Each kind requires its own aspect and especially its adaptation.

\paragraph{Singleton monitors}
In some cases it is sufficient to use a singleton instance of a monitor. This is the case when monitoring all the method calls that occur in a bundle, within all objects of a class, or within a singleton object. For monitoring method call orders, we use a {\tt before} pointcut in AspectJ. Figure~\ref{fig:exaspect} shows an example aspect: Here, before the calls to methods -- specified in the execution pattern after the ``:'' in the pointcut -- of all objects of class {\tt MiddlewareProc} an update on the state transition function -- the {\tt com.nextState} -- is inserted. We extract the name of the called method using reflection and a helper method {\tt AJMonHelpers.getMethodName} and pass it to the state transition function. In addition to updating the state field in the monitor we get a boolean value indicating whether the monitored property is still fulfilled. In case of a deviation the {\tt BehavioralTypeViolationException} -- a runtime exception is thrown. The implementation of the {\tt MiddlewareProc} class may or may not catch this exception and react to it.
\begin{figure}
\centering
{\small
\begin{verbatim}
package bookingsystem.middleware;

import java.util.HashMap;
import java.util.Map;
....
import monitors.*;


public aspect CallincprotocolMiddlewareProc {

    ...
	
    pointcut myMethod(MiddlewareProc p): this(p) && 
       within(MiddlewareProc) && execution(* *(..));
    before (MiddlewareProc p): myMethod(p) {

        ...

    	boolean verdict = com.nextState(
           AJMonHelpers.getMethodName(
              thisJoinPointStaticPart.getSignature()));
    	if (!verdict) throw new BehavioralTypeViolationException();       
    }
}
\end{verbatim} 
}

\caption{Example aspect}
\label{fig:exaspect}
\end{figure}

\paragraph{Multiple monitor instances}
In same cases we want to monitor each object of a class with an independent monitor. Here, we create on call of the object's constructor an individual monitor for the object. It is added to a (hash)map (Object $\rightarrow$ Monitor). Since the AspectJ pointcuts are defined with respect to the static control flow information specified in the source code of a class, on each call of a method belonging to the class to be monitored, we use the same code in each object and chose the  monitor for the particular object by looking it up from the map and advance the respective monitor state.

\paragraph{Monitoring of time}
Monitoring time is done using Java timers within the Java code associated with the pointcuts. On call of a method we create a timer that is scheduled to throw an exception after the specified maximal execution time. Using the {\tt after} pointcut, the timer is canceled if the method's execution finishes on time and thus, no exception is thrown in this case.

 The adaptation of an aspect for monitoring a particular component is simple. One has to take the appropriate AspectJ .aj file and adapt it, by inserting the names of the classes and packages that shall be monitored and the correct monitor names. Weaving of the aspects is done automatically on Java bytecode level and no additional configuration needs to be done.

\subsection{Simple Behavioral Type Checking}
\label{sec:deccomp}
We have developed and implemented different operations for handling and comparing behavioral types, for deciding compatibility and for deadlock freedom.

Simple comparison for equality of types and comparison for refinement between two automata based specifications involves the following steps.
\begin{itemize}
\item A basis for the comparison of two types is the establishment of a set of semantical artifacts (e.g., method calls) that shall be considered. The default is to use the union of all semantical artifacts that are used in the two types. Comparison for refinement is achieved by eliminating certain semantical artifacts from this set. For consistency this also requires eliminating associated transitions from the types  or, depending on the desired semantics, replacing an edge with an empty or $\tau$ label.
\item It is convenient to complete specifications for further comparison: Specification writer may only have specified method calls or other semantical artifacts that trigger a state change. Here, we automatically add an error location. We collect possible labels and for locations that do not have an edge for a label leading to another location indicating a possible semantical artifact, we add edges with the missing label to the error location.
\item In case of specifications which have been completed and that have no locations with two outgoing edges with the same labels, we perform a minimization of automata based specifications. This way, we merge locations and get rid of unnecessary complexity automatically.
\item Normalization of automata based specifications. This, involves the ordering of edges and in some cases locations with respect to the lexicographic order of their labels / location names.  
\item Checking for equality involves the checking of equality of the labels on edges. Optionally, one can also consider the equality of location names of an automaton. Location names may imply some semantics but in our standard settings they only serve as ids. When location names serve only as ids, we  construct a mapping between location names of the two automata involved in the comparison operation.
\end{itemize}
These operations have been implemented in Java.  They do not need additional tools or non-standard plugins.

\subsection{Deciding Compatibility and Deadlock-Freedom }

In addition to the operations described in Section~\ref{sec:deccomp} we have adapted a SAT and game-based tool -- VissBIP  presented in \cite{cheng2011synthesis} -- to serve as a compatibility and deadlock checker for our behavioral types for OSGi.
Our framework uses VissBIP to support the checking of the following properties:
\begin{itemize}
\item Deadlocks checking: deadlocks resulting from potential sequences of method calls can be detected.
\item Compatibility: A component anticipating a certain behavior of incoming method calls matches potential behavior of outgoing method calls by other components.
\end{itemize}

VissBIP uses a simplified version of the BIP semantics \cite{bip1}. A system comprises concurrent automata with labeled edges. The automata synchronize with each other by performing edges with the same labels in parallel.
Otherwise, the default case is that automata do not synchronize with each other. 
For comparing method call based behavioral specifications we use VissBIP  on specifications that comprise expected incoming and outgoing method calls of components. In OSGi synchronization between components happens only when one component calls a method of the other component as indicated in the behavioral specification and the OSGi semantics. On the VissBIP side this corresponds to same labels in the automata that represent the behavior.
In addition to the label compatibility checking, VissBIP is able to perform the introduction of priorities.

\subsection{Runtime Adaption of Systems}
One way of runtime adaption is the reaction to potential deadlocks or incompatibilities.
Recall Figure~\ref{fig:oldnewprot}: it shows behavioral specifications of two components which intend to communicate with each other. Possible outgoing method calls of one component and expected incoming method calls of the other component are shown. It can be seen that the first component is able to communicate using two different protocols: one starts by calling an initialization method {\sf newPrtcl}, the other one starts by calling an initialization method {\sf oldPrtcl}. The other component expects the {\sf newPrtcl} call. 

When we give these two specifications to VissBIP, it will return a list of priorities where the {\sf newPrtcl} edge is favored over the {\sf oldPrtcl} edge in the first specification. In a Java implementation the first component can use this to dynamically decide at runtime which protocol to use. 
\begin{itemize}
\item 
First, the component loads its own behavioral specification and the specification of the expected method calls of the second component. Technically, we support loading files and the registration of models as properties / attributes of bundles as provided by the OSGi framework.
\item 
Next, we invoke VissBIP or another checking routine. Passing the behavioral specifications as parameters.
\item
The checking routine gives us a list of priorities. In the Java code we have a switch statement as a starting point for handling the different protocols. We check the priorities and go to the case for the appropriate protocol.
\end{itemize}
Thus, in addition to deadlock detection, we can use behavioral specifications for coping with different versions of components and desired interacting protocols.

\subsection{Component Discovery at Runtime}
A central feature of our behavioral descriptions for OSGi components is registering them to a central OSGi instance. In order to inform other components of the existence of a bundle with behavioral offers and needs, we register its behavioral properties using the OSGi service registry belonging to a {\tt BundleContext} which is accessible for all bundles in the OSGi system:
\begin{verbatim}
  registerService(java.lang.String[] clazzes,
     java.lang.Object service,
     java.util.Dictionary<java.lang.String,?> properties)
\end{verbatim}
Here, we register a collection of behavioral objects as properties for a service representing a bundle under a String based key. In our framework, we register a collection of behavioral models as "BEHAVIOR". The behavioral models are loaded from XML files that are integrated into the bundle. The behavioral models come with meta information which identify the parts of the behavior of a bundle which they describe. The service itself is represented as an object. Additional interface information is passed using the {\tt clazzes} argument.

\section{Evaluation}
\label{sec:eval}

Our evaluation features a booking system (\cite{osgigeneraltech}) as an example. It is evaluated with respect to different aspects of behavioral types including behavioral runtime monitoring.

\subsection{Booking System}

\label{sec:booksys}
We present the use of behavioral types to highlight some features and usages of our work on an example: a flight booking system.

\begin{figure}
\centering
\includegraphics[width=0.5\textwidth]{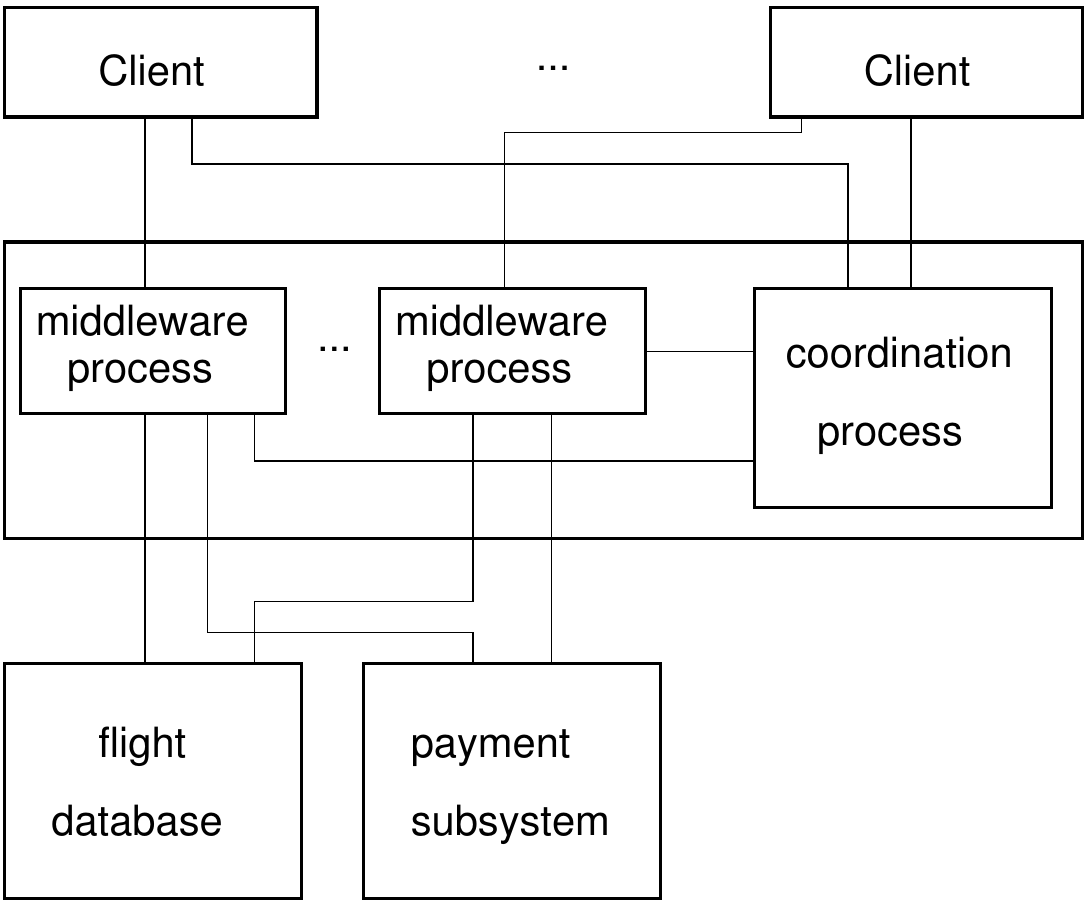}
\caption{Components of our flight booking system}
\label{fig:fbcomps}
\end{figure}

Figure~\ref{fig:fbcomps} shows the main ingredients of our flight booking system. Clients are served by middleware processes which are created and managed by a coordination process. Middleware processes use concurrently a flight database and a payment system. The described system is an example inspired by realistic systems where the middleware is implemented using Java/OSGi. In addition to the middleware components we describe databases and parts of the frontend using our behavioral types to make checks of these parts possible. 

\subsection{Behavioral Types for the Booking System}

The following means of behavioral interaction can be distinguished:
\begin{itemize}
\item {\bf Component calls between methods / communication protocol} In our flight booking system, a client can call a coordination process and middleware processes. Middleware processes can call methods providing access to the flight database and the payment subsystem. The method calls need to respect a distinct protocol which can be encoded using our behavioral types.
\item {\bf Creation and deletion of new components} The coordination process creates and removes middleware process such that there is one process per client. Providing support for analysis of such dynamic aspects is a long term goal for our behavioral types but not in the scope of this work.
\item {\bf Concurrent access to shared resources}
Middleware processes perform reservations, cancellations, rebookings, seat reservations and related operations on the flight database. These operations do require the locking of parts of the data while an operation is performed. For example, during a seat reservation a certain amount of the available seats in an aircraft is locked so that a customer can chose one without having to fear that another customer will chose the same seat at the very same time. In the current state we are able to provide some behavioral types support here.
\end{itemize}

\begin{figure}
\centering
\includegraphics[width=0.7\textwidth]{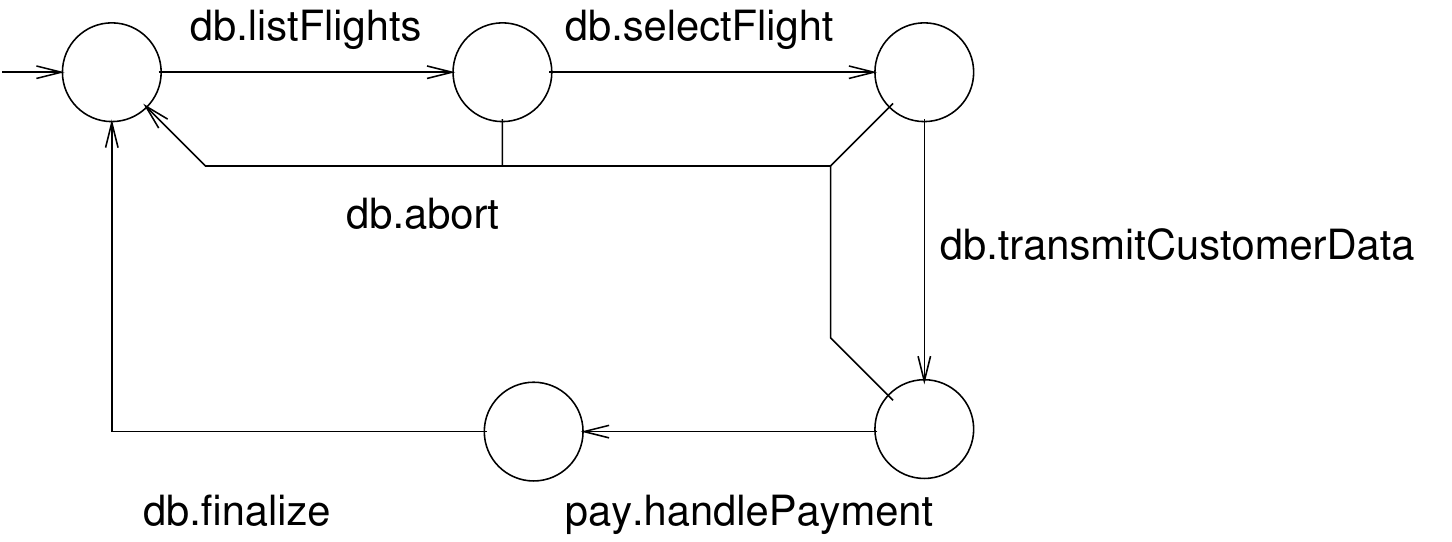}
\caption{Outgoing method calls of a middleware process}
\label{fig:commprot1}
\end{figure}

\paragraph{Example: Specification of outgoing method calls of a middleware process}
Specifications of possible expected incoming and potential outgoing method calls give information about a communication protocol that is to be preserved. Typically different interaction sequences are possible, especially since we are dealing with abstractions of behavior.
In the booking system, a middleware process communicates with a flightdatabase (db) and the payment system (pay). The expected order of method calls for a flight booking to these systems is shown in Figure~\ref{fig:commprot1}.
The figure shows only an excerpt of the possible states and transitions. In addition to this, the initial state allows the start of a seat reservation process and a cancellation process. Moreover,
Figure~\ref{fig:commprot1} shows only the state changing method calls of the behavioral specification of the booking process. Our real behavioral specification completely lists all possible method calls in each state. This way, we can further analyze compatibility issues for example with database systems that do not support all possible method calls of a middleware process.  

In comparison to the outgoing method calls of a middleware process, the incoming method call specification is much simpler: A constructor call is performed by the coordination process upon initialization. After that, the communication with the client is done using a webserver interface -- comprising method calls that send raw request data to the middleware process and return raw response data that trigger, e.g., displaying selected flights by the client -- where no states in the communication process can be distinguished. 

\paragraph{Example: Specification of database elements}
Access to our database is done using method calls to a database process and is formalized using our automata based specification formalisms. The method calls result in locking and unlocking database elements. Seat reservation in a flight requires that a certain partition of the available seats is blocked during the selection process so that a client can make a choice.

\begin{figure}
\centering
\includegraphics[width=0.75\textwidth]{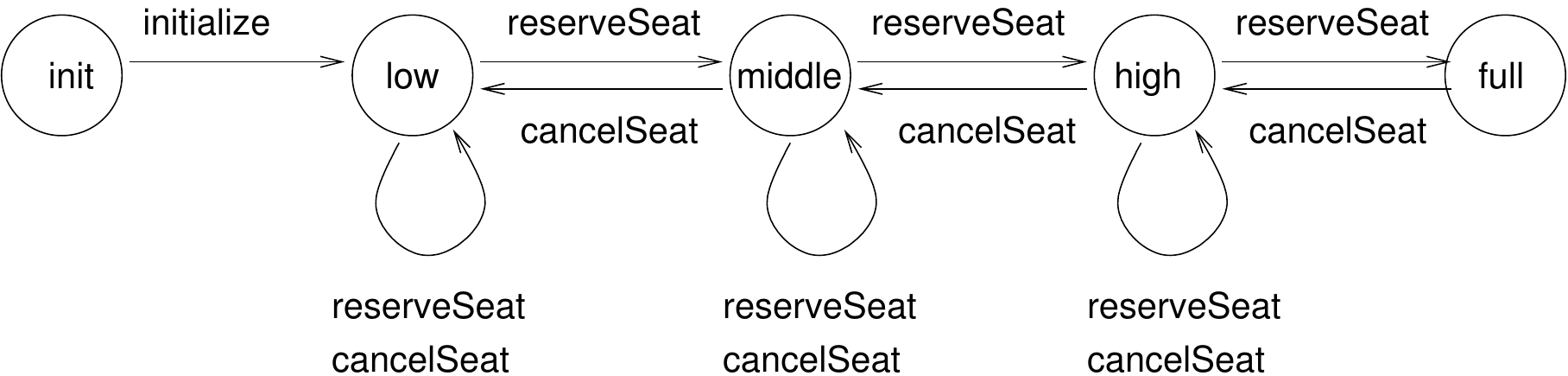}
\caption{Behavioral model for seat reservation of a flight}
\label{fig:seatres}
\end{figure}

Figure~\ref{fig:seatres} shows our behavioral model of seat reservation for a single flight. Different loads are distinguished: low means that many seats are still available, while high means that only a few seats are available. The full state indicates that no additional seat reservations can be made, only cancellations are possible. The model is an abstraction of the reality since instead of treating each seat -- potentially hundreds of available seats -- independently we only distinguish their partitioning into four equivalence classes: low, medium, high and full.


\paragraph{Example: Database elements and deadlocks}
Access to the flight database can result in deadlocks. The model from Figure~\ref{fig:seatres} can serve as a basis for deadlock analysis. Consider the scenario shown in Figure~\ref{fig:concseatres}:
\begin{figure}
\centering
\includegraphics[width=0.45\textwidth]{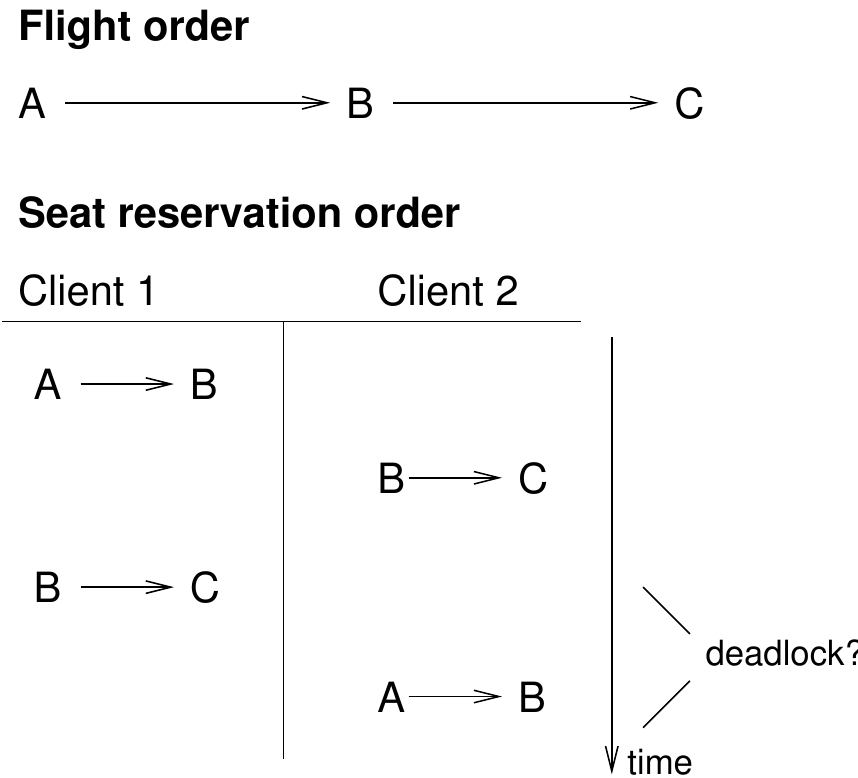}
\caption{Concurrent seat reservation on two flights}
\label{fig:concseatres}
\end{figure}
For each flight a different instance of the seat reservation model exists. Given three airports A, B and C: Suppose two people -- person 1 and person 2 -- want to fly from A to C via B. Seats for two flights need to be reserved: from A to B and from B to C. It is not desirable to reserve a seat from B to C if no seat is available for the flight to A to B. Otherwise, it might not be desirable to fly from A to B if no seat is available for the flight from B to C. 

During the seat reservation a deadlock may occur: If person 1 reserves the last seat for the A to B flight before doing reservations for the B to C flight and person 2 reserves the last seat for the B to C flight before a seat reservation for the A to B flight a deadlock may occur, which may result in the cancellation of both journeys although one person could have taken the journey.

If it is known before to the seat reservation system that person 1 and person 2 will fly from A to C -- which is a reasonable assumption given the fact that they have entered their desired start and end destination into the system --  we are able to detect such deadlocks. They can occur if both behavioral models of the seat reservation system are already in the high state -- given that no other participants are doing reservations at this time we may also take compensating actions. 

\subsection{Behavioral Runtime Monitoring and the Booking System}
Different scenarios for the use and deployment of behavioral types have been tested by us. One example scenario is the flight booking system. OSGi components and their interactions are shown.  The entire system could be deployed as an OSGi based middleware that offers its services to the external world using webservices. Clients are represented as proxy components in the system and served by middleware processes which are created and managed by a coordination process. Middleware processes use concurrently a flight database and a payment system which are represented by proxy OSGi components.  We have investigated the communication structure between the components and investigated deployment of monitors. This  comprises the following cases:
\begin{itemize}
\item
The use of multiple monitors running in parallel and being created at runtime for different objects which are created dynamically. In the example system this is the case for the middleware processes, where processes are created as separate objects on demand and  are monitored independently of each other. 
\item
The monitoring of all objects of a single class using a single monitor and the monitoring of singleton objects and the monitoring of bundle behavior. This is, e.g., the case in the payment subsystem. 
\end{itemize}
Different Aspects as described in Section~\ref{sec:aspects} were adapted for this. Monitors together with an implementation of the system that realized the communication between the components was deployed using the OSGi Equinox implementation \footnote{Making AspectJ and OSGi run together can require some extra work. A solution for Equinox is described at \url{http://eclipse.org/equinox/weaving/}}. Furthermore, we have investigated the monitoring of maximal execution time of methods. In the example system this is the case in the payment subsystem and access to the flight database. We did not find any major problems in our approach.

\paragraph{Behavioral Runtime Monitors and OSGi}
As described in Section~\ref{sec:beht} our behavioral type efforts are particularly aimed towards OSGi. While features that are not subject to the contribution of this paper like component discovery using behavioral types are only feasible in an OSGi like component framework, we did find no principal issues that prevent the use of our types for behavioral runtime monitoring in other Java contexts. Technically, the OSGi framework gives us with bundles yet another structuring layer for software components, which we use in the specification of our types and we believe has a good granularity for the communication protocol specifications that we primarily regard. The generation of entire classes for each monitor instead of integrating the complete monitor inside of aspects like in \cite{mop} is also justified on this granularity. The editors and generation mechanism depend on Eclipse which is realized on-top of OSGi. The behavioral monitor connection using aspects depends on AspectJ, some technical issues are mentioned above.
\section{Related Work}
\label{sec:rw}
Interface automata \cite{AlfaroHenzinger:2001} are one form of behavioral types. Like in this work, component descriptions are based on automata. The focus is on communication protocols between components which is one aspect that we also address in this paper. While the used formalism for expressing behavior in interface automata is more powerful (timed automata vs. automata vs. timing annotation per method), interface automata do not target the main focus of this paper: checking the behavior at runtime of a component by using some form of monitoring. They are especially aimed at compatibility checks of different components interacting at compile time of a system.
The term behavioral types  is used in the Ptolemy framework \cite{betypesptolemy}. Here, the focus is on real-time systems.

Specification and contract languages for component based systems have been studied in the context of web services. A process algebra like language and deductive techniques are studied in \cite{webcontracts}. Another process algebra based contract language for web services is studied in \cite{webcontrmath}. Emphasize in the formalism is put on compliance, a correctness guaranty for properties like deadlock and livelock freedom. Another algebraic approach to service composition is featured in \cite{conssercomp}.

JML \cite{jml} provides assertions, pre- and postconditions for Java programs. It can be used to specify aspects of behavior for Java methods.  
Assertion like behavioral specifications have also been studied in the context of access permissions~\cite{plural}.

Behavioral types as means for behavioral checks at runtime for component based systems have been investigated in~\cite{abt}. In this work, the focus is rather put on the definition of a suitable formal representation to express types and investigate their methodical application in the context of a model-based development process. 

A language for behavioral specification of components, in particular of  object oriented systems -- but not OSGi --, is introduced in~\cite{brochjohnsen12}. Compared to the requirement-based descriptions proposed in our paper, the specifications used in \cite{brochjohnsen12}  are still relatively close to an implementation. Recent work regarding refinement of automata based specifications is, e.g., studied in \cite{behavioralrefinement}.

The runtime verification community has developed frameworks which can be used for similar purpose as our behavioral type based monitors. The MOP framework \cite{mop} allows the integration of specifications into Java source code files and generates AspectJ aspects which encapsulate monitors.  Compared to this work, the intended goals are different. While we keep the specification and implementation part separate, in order to be able to use the specification for different purposes at development, compile and runtime, a close integration of specification and code is often desired and realized in the runtime verification frameworks.  A framework taking advantage of the trade-off between checking specifications at runtime and at development time has been studied in \cite{bodden12}. A framework that generates independent Java monitors leaving the instrumentation aspect to the implementation is described in \cite{barringer04}. Other topics explored in this context comprise, e.g.,  the efficiency and expressiveness of monitoring \cite{BFHRR12,BauerL11} but are less focused on software engineering aspects compared to this paper.

Monitoring of performance and availability attributes of OSGi systems has been studied in \cite{osgiperfmon}. Here, a focus is on the dynamic reconfiguration ability of OSGi. Another work using the .Net framework for runtime monitor integration is described in \cite{sch06}. 
Runtime monitors for interface specifications of web-service in the context of a concrete e-commerce service have been studied in~\cite{halleetal2010}. Behavioral conformance of web-services and corresponding runtime verification has also been investigated in~\cite{cao10}. Runtime monitoring for web-services where runtime monitors are derived from UML diagrams is studied in~\cite{gan2007runtime}.

Runtime enforcement of safety properties was initiated with security automata~\cite{secpol} that are able to halt the underlying program upon a deviation from the expected behaviors. In our behavioral types framework, the enforcement of specifications is in parts left to the system developer, who may or may not take potential Java exceptions resulting from behavioral type violations into account.

Our behavioral types represent an abstract view on the semantics of OSGi. We have summarized our work on the OSGi semantics in a report \cite{reportosgisem}. Other work does describe OSGi and its semantics only at a very high level. A specification based on process algebras is featured in \cite{mekontsotchinda:inria-00619233}. Means for ensuring OSGi compatibility of bundles realized by using  an advanced versioning system for OSGi bundles based on their type information is studied in \cite{brada09}.
Some investigations on the relation between OSGi and some more formal component models have been done in \cite{muellerse2010}. Aspects on formal security models for OSGi have been studied in \cite{osgisecuritybycontr}.

\section{Conclusion}
\label{sec:concl}

We presented our BehT framework for behavioral types for OSGi systems, a development process for OSGi applications and some motivation and evaluation. 

So far, we are concentrating on Eclipse / OSGi systems. Other application areas for the future comprise 1) work towards behavioral types for distributed software services 2) work towards real-time embedded systems. This might require leaving the Java / OSGi setting, since these applications typically involve C code which communicates directly with -- if at all -- an operating system. There is, however, work on extensions for real-time applications of OSGi using real-time Java (e.g., \cite{donsez12}). Additional specification formalisms and the integration of new checking techniques are another challenge.

\bibliographystyle{plain}

\end{document}